\documentclass[aps,prb,showpacs,twocolumn,amsmath,amssymb]{revtex4}
\usepackage[dvips]{color,graphics,epsfig,rotating}
\usepackage{graphicx}% Include figure files
\usepackage{dcolumn}% Align table columns on decimal point
\usepackage{bm}% bold math

\begin{document}

\title{First-principles investigation of effect of pressure on BaFe$_2$As$_2$}

\author{Wenhui Xie, Mingli Bao, Zhenjie Zhao}
\affiliation{Department of Physics, East China Normal University,
Shanghai, 200062, China}\affiliation{Engineering Research Center
for Nanophotonics and Advanced Instrument, East China Normal
University, Shanghai, 200062, China}

\author{Bang-Gui Liu}
\affiliation{Institute of Physics, Chinese Academy of Sciences,
Beijing 100190, China}

\date{\today}

\begin{abstract}
On experimental side, BaFe$_2$As$_2$ without doping has been made
superconducting by applying appropriate pressure (2-6 GPa). Here, we
use a full-potential linear augmented plane wave method within the
density-functional theory to investigate the effect of pressure on
its crystal structure, magnetic order, and electronic structure. Our
calculations show that the striped antiferromagnetic order observed
in experiment is stable against pressure up to 13 GPa. Calculated
antiferromagnetic lattice parameters are in good agreements with
experimental data, while calculations with nonmagnetic state
underestimate Fe-As bond length and c-axis lattice constant. The
effects of pressure on crystal structure and electronic structure
are investigated for both the antiferromagnetic state and the
nonmagnetic one. We find that the compressibility of the
antiferromagnetic state is quite isotropic up to about 6.4 GPa. With
increasing pressure, the FeAs$_4$ tetrahedra is hardly distorted. We
observe a transition of Fermi surface topology in the striped
antiferromagnetic state when the compression of volume is beyond 8\%
(or pressure 6 GPa), which corresponds to a large change of $c/a$
ratio. These first-principles results should be useful to
understanding the antiferromagnetism and electronic states in the
FeAs-based materials, and may have some useful implications to the
superconductivity.
\end{abstract}


\maketitle

\section{introduction}

The discovery of high temperature superconductivity in
iron-pnictide materials\cite{kamihara} has attracted widespread
interest in elucidating the superconductivity mechanism and
searching for new high-T$_c$
materials\cite{lafeaso:pickett,mazin,singh-du}. The highest
critical temperature ($T_c$) observed in the RFeAsO series
compounds (R= La-Gd) is up to 55 K with electron
doping\cite{liu,cwang}. Recently, superconductivity was discovered
in the AFe$_2$As$_2$ materials (A = Ca, Ba, Sr) which have similar
FeAs layers. It has been reported that the transition temperature
can be as high as 38 K for (K,Ba)Fe$_2$As$_2$ and
(K,Sr)Fe$_2$As$_2$\cite{bakfeas:38K,bakfeas:lat}. All of these
compounds show an  antiferromagnetic order and a structural
transition with similar ordering temperatures\cite{lafeas:SDW}.
Although the superconductivity mechanism is not yet known up to
now, the phonon mediated pairing seems to be ruled
out\cite{boeri:phonon}. The fact that antiferromagnetic spin
fluctuations coexist with superconductivity indicates that the
nature of superconductivity should be complex and unconventional
in the FeAs-based
compounds\cite{lafeaso:pickett,mazin:dftsum,singh-du}, like that
of the cuprate high-T$_c$ superconductors.

Stoichiometric BaFe$_2$As$_2$ is not superconductor but an
antiferromagnet\cite{bafeas:SDW,bafeas:magstr1}. Superconductivity
can be achieved by either electron or hole doping. In hole doped
superconducting (K,Ba)Fe$_2$As$_2$, $T_c$=38K has been obtained by
replacing 40\% Ba by K\cite{bakfeas:38K}. So that BaFe$_2$As$_2$
is a typical parent compound of the high temperature
iron-arsenides superconductors. In addition, recent experiments
confirmed that BaFe$_2$As$_2$ without doping becomes
superconducting under high pressure\cite{bafeas:28K}. The pressure
inducing superconductivity was also observed in the same group
compounds SrFe$_2$As$_2$\cite{bafeas:28K} and
CaFe$_2$As$_2$\cite{cafeas:sup}. It is very interesting to
investigate the properties of AFe$_2$As$_2$  materials under high
pressure.

In this paper we use full-potential augmented-plane-wave method
based on the density-functional theory to investigate pressure
effect on the crystal structure, magnetism and electronic structure
of BaFe$_2$As$_2$. The reason we concentrate on BaFe$_2$As$_2$ to
investigate the pressure effect is that its superconductivity, in a
wide pressure range (between 2-6 GPa) and the highest $T_c$ up to 28
K, is more robust than that of the Sr and Ca
compounds\cite{bafeas:28K}. At first, we analyze the effects of
different magnetic orders on the equilibrium lattice parameters and
corresponding electronic structures. Our calculation indicates that
the striped antiferromagnetic state is the ground state, its
calculated crystal structure and lattice constants are consistent
with experimental data. With increasing pressure, there is a
topological transition of Fermi surfaces in the striped
antiferromagnetic state approximately at 6 GPa. This Fermi surface
transition is possibly related to the structural transition and
consistent with the loss of superconductivity at the pressure. These
full-potential DFT results should be helpful to understand the
pressure-induced properties of BaFe$_2$As$_2$\cite{mazin:dftsum}.

The remaining part of this paper is organized as follows. In next
section, we describe our computational method and parameters. In
section III, we analyze the crystal structure, magnetic orders and
electronic structures under ambient pressure. In Section IV, we
present pressure-driven changes of the crystal structure, magnetic
order, and electronic structures. In section V, we compare our
calculated results with known experimental results. Finally, we
give our conclusion in section VI.

\section{Computational method}

The present calculations are performed by using an {\it ab initio}
all-electron full-potential linearized augmented plane wave (FLAPW)
method based on  the density-functional theory, as implemented in
WIEN2K code \cite{wien2k}. Both local density approximation (LDA)
\cite{lda92} and generalized gradient approximation (GGA)
\cite{pbe96} are used in our calculations. The atomic sphere radii
of 2.4, 2.0, and 1.9 atomic unit are used for Ba, Fe and As,
respectively. The self-consistent calculations are considered to be
converged only when the integrated charge difference between input
and output charge density is less than 0.0001. The internal
parameter $z_{\rm As}$, describing the As position, is relaxed until
the force per atom is smaller than 1mRy/a.u. We use 500 $k$-points
for Brillouin-zone integrations in the antiferromagnetic
calculations, and at least 1500 $k$-points in the nonmagnetic
calculation.

BaFe$_2$As$_2$ has ThCr$_2$Si$_2$ type body centered tetragonal
structure ($I4/mmm$) with lattice parameters $a$ = 3.9625 \AA{}
and $b$ = 13.0168 \AA{} at room temperature. The FeAs layers are
stacked so that the As atoms face each other and the Ba atoms sit
in the resulting 8-fold coordinated square prismatic sites between
them. At 140 K, it undergoes a structural phase transition from
tetragonal to orthorhombic structure ($Fmmm$) with lattice
constants $a$ = 5.6146 \AA{}, $b$ = 5.5742 \AA{}, and $c$=12.9453
\AA{} (at 20K), accompanied by a magnetic phase transition from a
Pauli paramagnetism to an antiferromagnetic order. Neutron
scattering experiments have shown that the magnetic moments are
aligned ferromagnetically along $b$-axis but antiferromagnetically
along $a$-axis and $c$-axis\cite{bafeas:neutron,bafeas:magstr1},
the named striped antiferromagnetic order state (AF1). We also
calculated checkerboard antiferromagnetic order state (AF2), in
which magnetic moment of nearest-neighbor Fe are
antiferromagnetical order while magnetic moment of next
nearest-neighbor are ferromagnetical order. In all the
calculations, we keep $b$/$a$ ratio in term of the experimental
value because we have found that the changing of $b$/$a$ has tiny
influence on the total energy and electronic structure.

\section{analysis of ambient-pressure properties}

As a starting point, we took the experimental lattice parameters
of tetragonal paramagnetic state and the striped antiferromagnetic
state of BaFe$_2$As$_2$, and relaxed the internal coordinate of As
by LDA calculation with generalized gradient approximation (GGA).
For nonmagnetic state, we found that relaxed internal parameter of
$z_{\rm As}$=0.3448 is much lower than the experimental value of
$z_{\rm As}$=0.3545. While for the striped antiferromagnetic
order, we arranged the direction of iron magnetic moment in terms
of experimental observation\cite{bafeas:neutron}, and got a value
of $z_{\rm As}$=0.3515, which is in better agreement with
experimental data of $z_{\rm As}$=0.3538 \cite{bafeas:neutron}.
Other antiferromagnetic settings with optimized lattice parameters
cause smaller $z_{\rm As}$ value, but are better than nonmagnetic
result. Calculated magnetic moment of iron inside muffin tin
sphere is 1.86$\mu_{B}$, almost twice larger than the value of
0.8$\mu_{B}$ measured by neutron scattering
experiment\cite{bafeas:neutron}.
\begin{figure}
\includegraphics[width=8.6cm,angle=0]{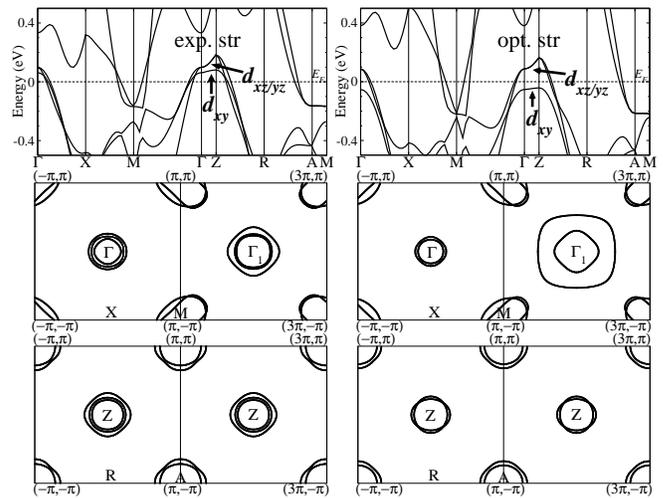}
\caption{\label{tetfs}  Band structure (top two panels) and Fermi
surface sections (the other four panels) of BaFe$_2$As$_2$ of
experimental lattice constant with experimental As coordinates
$z_{\rm As}$ (left column, labelled as exp. str) and with {\it ab
initio} optimized value (right column, labelled as opt. str). The
middle panels show the Fermi surface sections extended to the second
Brillouin zone in the $k_x$ direction for the $k_z$=0 plane, and the
bottom panels show those for the $k_z$=$\pi$/c plane.}
\end{figure}

It turns out that the band character of BaFe$_2$As$_2$ is similar to
that of LaFeAsO\cite{boeri:phonon}, except that La $f$ and O $p$
staes are missing in BaFe$_2$As$_2$ compound. The band splitting of
$d_{3z^2-1}$ states is also significantly weaker in BaFe$_2$As$_2$
than in LaFeAsO, which indicates that interlayer bonding through the
Ba layer of BaFe$_2$As$_2$ is weaker than the LaO layers of LaFeAsO.
Since there have been many reports on the electronic
structures\cite{bafeas:singhband,bafeas:xiangband,bafeas:fangband,bafeas:nekrasovband},
we show only the bands near Fermi level to manifest the key parts.
The band structure and Fermi surfaces of tetragonal phase are
illustrated in Fig. \ref{tetfs} for both experimental and optimized
structures.

Fermi surfaces of both experimental and optimized structures have a
doubly-degenerated electron pocket centered at the $M$ point. These
sheets have a dominant $d_{xz},d_{yz}$ character, only a little
component of $d_{xy}$ character. Around $\Gamma$ point, there is
significant difference. For optimized structure, Fermi surfaces
comprise a doubly-degenerated cylindrical hole pocket centered at
the $\Gamma$ point, mainly consisting of $d_{xz/yz}$ character and
little mixture of $d_{x^2-y^2}$ and $d_{3z^2-1}$. While for
experimental structure, another new cylindrical hole sheet, mainly
$d_{xy}$ character, inserted to the two hole sheets. Those sheets
have more two-dimensional character. Compared to the observation of
APRES experiment\cite{arpes:feng}, the calculated results using the
experimental lattice parameters are in better agreement with ARPES
data. Furthermore, in comparison with other calculated
results\cite{bafeas:xiangband,bafeas:fangband,bafeas:singhband,bafeas:nekrasovband},
we would find that whether $d_{xy}$ bands cross Fermi level or not,
it is sensitive to $z_{\rm As}$, which is similar to the results
discussed in LaFeAsO system \cite{mazin:dftsum,bafeas:singhband}.

%added by xiewh
Furthermore, it is found that the band width of the $d_{xy}$ bands
changes little, but they shift downwards by about 0.2 eV in
comparison to the $d_{xz/yz}$ bands. This fact indicates that the
tiny change of As position has a significant effect on the crystal
splitting between $d_{xy}$ and $d_{xz/yz}$ orbital bands. Since
there is direct $d_{xy}$ coupling between the two Fe atoms, the
$d_{xy}$ bands are not sensitive to the change of the $z_{As}$
position and are more likely affected by the Fe-Fe distance, while
$d_{xz/yz}$ orbital are likely coupled with As $p$ orbital states
(mainly $p_{x}$ and $p_{y}$), the $d_{xz/yz}$ bands have stronger
$k_z$ dispersion than the $d_{xy}$ bands and the corresponding Fermi
surfaces have more three-dimensional (3-D) character, which change
more distinctively as the Fe-As bond length is shortened.
The discrepancy of Fe-As bond length between {\it ab initio}
calculations and experimental data, which is induced by the
magnetism, is a common feature of superconducting iron-pnictide
family. This is because the iron-pnictide materials is in proximity
to a quantum critical point, results in a strong spin
fluctuations\cite{mazin:dftsum}.

%%%%%%%%%%%%%%%%% striped AFM band and FS
\begin{figure}
\includegraphics[width=7cm,angle=0]{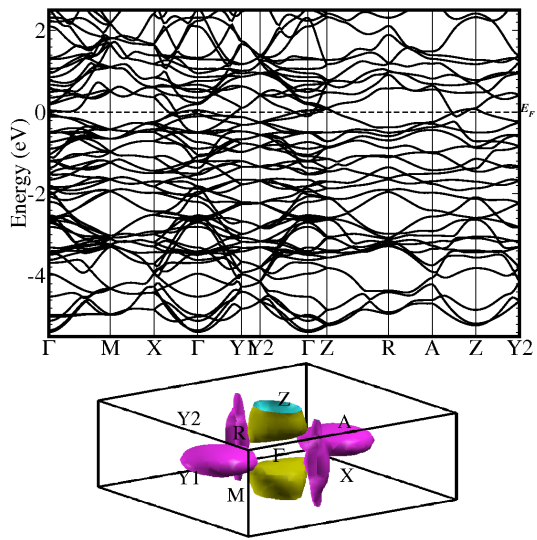}
\caption{\label{pv100} (color online). Band structures and Fermi
surfaces of the striped antiferromagnetic state with experimental
lattice. Here, we use a orthogonal lattice and show Fermi surfaces
on a conventional orthogonal BZ. The choice of BZ does not influence
our main discussions and results, but is better to guide for eyes. }
\end{figure}
%

%%% DOS added according to referee's sugguestion
%%%%%%%%%%%%%%%%% striped AFM band and FS
\begin{figure}
\includegraphics[width=7.5cm,angle=0]{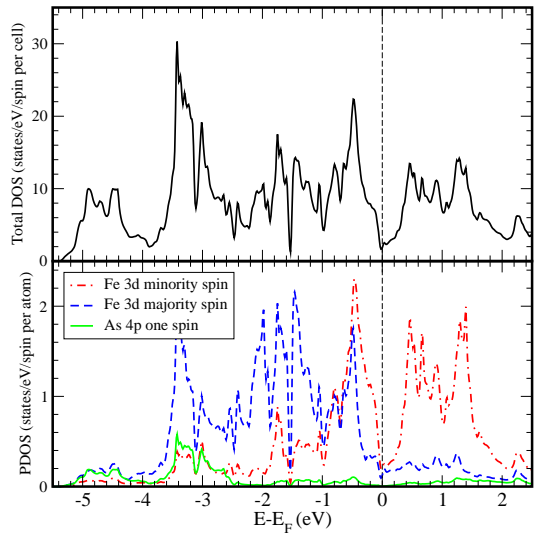}
\caption{\label{pdos} (color online). Top panel: total DOS for the
striped antiferromagnetic state. Bottom panel: spin resolved Fe 3$d$
DOS, showing majority (blue) and minority (red), and the As 4$p$ DOS
(green). }
\end{figure}

In Fig. \ref{pv100}, we show the band structures and Fermi surfaces
of the striped antiferromagnetic state with experimental lattice
constants.
%By xie 2009.01.19
In Fig. \ref{pdos}, we show the total density of states (DOS) and
the partial DOS of Fe 3$d$ and As 4$p$ states in different boxes.
The bands from $\sim-6$ eV to $-2$ eV are mainly from As $p$ states.
The Fe $d$ bands range from $-2$ eV to $+3$ eV, being above As $p$
states. Around $-3$ eV, there is significant hybridization between
the Fe $d$ states and As $p$ states. The band character around the
Fermi level is dominated by $d$ orbital. Due to the striped
antiferromagnetic order, a pseudogap is opened near the Fermi level
around $\Gamma$ point.
%added by xie 2009.01.19
In Fig. \ref{pv100}, we could find the shape of Fermi surfaces is
not regular due to the complex $d$ bands dispersion around the Fermi
level induced by antiferromagnetic interaction. There are hole-type
Fermi surfaces around Z point (yellow), two slender electron-type
Fermi surfaces parallel to $\Gamma$-Z line located at $k_x$
direction, and two irregular electron-type Fermi surfaces located at
$\Gamma$-Y1 line (pink) which are separated and confined in one
quarter of BZ. They are mainly of $d_{xy}$, $d_{xz}$ and $d_{yz}$
characters, with some mixing of $d_{x^2-y^2}$ and $d_{3z^2-1}$
characters.
%
%We would find Fermi surfaces are centered around Z point and new
%pockets appear along antinodal directions,

% added by xie
Compared to the quasi-two-dimensional Fermi surfaces of the striped
antiferromagnetic LaOFeAs\cite{lafeaso:pickett}, the Fermi surfaces
of BaFe$_2$As$_2$ have distinct three-dimensional topology. The
effect of the orthorhombic lattice distortion on the electronic band
structure as well as the Fe moments is very weak. The inter-layer
spin interaction is antiferromagnetic because our total energy
calculations show that the ferromagnetic order state is higher by
about 2.6 meV in total energy per Fe atom. The inter-layer spin
coupling constant is estimated to be $J_{\perp}$ = 0.8 meV. The most
stable striped antiferromagnetic order state with a magnetic moment
1.86$\mu_{B}$ is energy lower 124 meV per Fe atom than the
nonmagnetic state, while the checkerboard antiferromagnetic order
state with a magnetic moment 1.63$\mu_{B}$ is only 54 meV energy
lower. By mapping to a simple Heisenberg model only including the
1st and 2nd neighbor exchange couplings, we could estimate the
coupling constants: $J_{1a}$ = 28.2 meV, $J_{1b}$ = 27.3 meV and
$J_2$ = 17.4 meV. The tiny difference between the coupling constants
$J_{1a}$ (along $a$-axis) and $J_{1b}$ (along $b$-axis) indicates
that exchange coupling of BaFe$_2$As$_2$ is not so anisotropic.

\section{effect of pressure on crystal structure, magnetic order, and electronic structure}

\begin{figure}
\includegraphics[width=3.2in,angle=0]{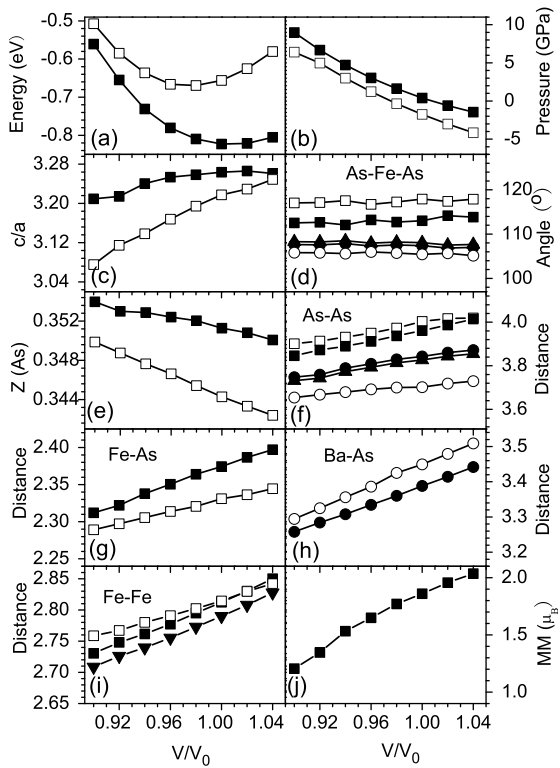}
\caption{\label{sum} Calculated values of total energy (a), pressure
(b), $c/a$ ratio (c), As-Fe-As angle (d), $z_{\rm As}$ (e), Fe-As
bond length (f), As-As bond length (g), Ba-As bond length (h), Fe-Fe
bond length (i), magnetic moment of iron (j) as functions of volume
($V_{0}$ is the experimental volume of the antiferromagnetic state
per formula unit, which is about 0.2\% smaller than the experimental
value of nonmagnetic state). Filled symbols indicate the striped
antiferromagnetic results, and open symbols nonmagnetic ones. For
the As-As distance and As-Fe-As angle, due to the orthorhombic
distortion, antiferromagnetic state has three values while
nonmagnetic state has only two.}
\end{figure}

Now we investigate the  effect of pressure on the crystal
structure and electronic structure of BaFe$_2$As$_2$. We compress
crystal volume up to 18\%, corresponding to pressure of 18 GPa
(about 180 kbar). We relax the internal parameters $z_{\rm As}$
and $c/a$ ratio (keeping $b/a$ ratio of orthorhombic structure in
term of the experimental value) of all structures at intervals of
2\% volume compression. We subsequently fit the E$\sim$V curves to
a Birch-Murnagham equation-of-state, to obtain the equilibrium
volume V$_0$ and the bulk modulus $B_0$ for each system. The
magnetic moment decreases with increasing pressure, which results
in possible small change of $b$/$a$, but this change hardly
influences other structural parameters such as $z_{\rm As}$ and
the electronic structure. Therefore, it is reasonable to keep
$b$/$a$ in our magnetic calculations.

The theoretical lattice constants and internal parameters obtained
by the complete optimization are listed in Table \ref{tab-struct}.
As expected, nonmagnetic calculations underestimate $c/a$ ratio
because As atom is more close to Fe layers. We could find that
nonmagnetic calculation roughly predict the in-plane Fe-Fe bonding
correctly, but shorten out-of-plane Fe-As bond length, therefore
As atom is more close to iron layer, resulting in a small $z_{\rm
As}$. While the calculated lattice parameters of the striped
antiferromagnetic state are in better agreement with experimental
data. Furthermore, the striped antiferromagnetic calculation could
also describe out-of-plane Ba-As and As-As bond better.

\begin{table}[tbp]
\caption{Calculated equilibrium lattice constant (in unit of
\AA{})with internal parameter ($z_{\rm As}$) of BaFe$_2$As$_2$
obtained from the fully optimized nonmagnetic (NM) and striped
antiferromagnetic (AF1) calculations.} \label{tab-struct}
\begin{tabular}{lcccc}
\hline
      &    exp.(NM)  & exp.(AF1) &   cal.(NM)  & cal.(AF1) \\
\hline
Spacegroup      &  $I4/mmm$  &  $Fmmm$    &  $I4/mmm$  &   $Fmmm$    \\
$a$(\AA{})        &   3.9625   &  5.6146    &   3.9591   &   5.6378    \\
$b$(\AA{})        &            &  5.5742    &            &   5.5931    \\
$c$(\AA{})        &   13.0168  & 12.9453    &   12.6210  &  12.9624    \\
$z_{\rm As}$    &   0.3545   &  0.3538    &   0.3448   &   0.3512    \\
$d$(As-Fe)(\AA{}) &   2.403    &   2.392    &   2.3330   &   2.3789    \\
\hline
\end{tabular}
\end{table}

In Fig. \ref{sum}, we show the dependence of structural parameters
on the compression of volume for both nonmagnetic and striped
antiferromagnetic states. The total energy, pressure, the bond
length, bond angle, and the magnetic moment of iron are illustrated
as functions of the compression of volume. The calculations indicate
that the striped antiferromagnetic state is the ground state as long
as the pressure is less than 15 GPa, about 16\% compression of
volume, then the spin-polarization disappears. It is found that
ferromagnetic state is unstable under a large range of pressure. The
checkerboard antiferromagnetic order is always higher in total
energy due to the smaller calculated magnetic moment, which results
in smaller magnetic exchange energy. It also could not give
reasonable lattice parameters, for example, to predict smaller $c/a$
ratio and $z_{\rm As}$.
% added by xie
As the volume is compressed by 8\%, the corresponding magnetic
moment of the checkerboard antiferromagnetic order state decrease
from 1.6 $\mu_{B}$ to 0.5 $\mu_{B}$. After having tried several
other antiferromagnetic configurations, we find that the striped
antiferromagnetic state is the most stable due to its largest
magnetic moment. Some of other configurations even could not
converge to magnetic solutions.

The high pressure experiment indicates that superconductivity
appears at about 2.5 GPa, then $T_c$ reaches a maximum near 3.5 GPa,
and disappears above 5.6 GPa\cite{bafeas:28K}. As shown in Fig.
\ref{sum}(b), the compression of volume could be estimated to cover
such a range of pressure. In terms of the antiferromagnetic results,
as volume is compressed from 2\% to 8\%, the pressure range relevant
to superconducting dome is covered completely, while in the
nonmagnetic calculation, the volume needs to be compressed more,
from 4\% to 10\%.

It is clear that Fe-As bond length is enhanced significantly than
Fe-Fe after considering the spin-polarization. Hence, As atom goes
away from Fe layer giving a larger $z_{\rm As}$. With a 10\%
compression of volume, the magnetic moment of iron deceases from 1.8
$\mu_B$ to 1.2 $\mu_B$, which reduces the discrepancy of Fe-As bond
length between nonmagnetic and antiferromagnetic states. In contrast
to tendency of iron related bonding, Ba-As  bond length becomes
shorter after considering the spin-polarization. In the striped
antiferromagnetic state, chemical bonding becomes more rigid against
pressure, therefore $c/a$ ratio changes less than nonmagnetic state,
as a result, the compressibility is less anisotropic.

Although the Fe-Fe bond length and Fe-As bond length decrease
significantly with the increase of pressure, the As-Fe-As bond
angles $\varepsilon$ change slightly. Under 10\% compression of
volume, three As-Fe-As angles of antiferromagnetic lattice are found
to be 107.6$^\circ$, 108.3$^\circ$ and 112.5$^\circ$, roughly
changed one degree compared to the ambient condition. Former two
values correspond to the As-Fe-As angles whose two As atoms
separated by the iron plane, and the latter one represents two As
atoms on the same side. Due to the stronger Fe-As bonding in the
antiferromagnetic state, the FeAs$_4$ tetrahedra gets more elongated
along $c$-axis, so that FeAs$_4$ tetrahedra in the antiferromagnetic
state is distorted less than that in nonmagnetic state, but it is
still far away from the ideal structure.

%%%%%%%%%%%%%%%%% comparing the BANDS and FS
\begin{figure}
\includegraphics[width=8cm,angle=0]{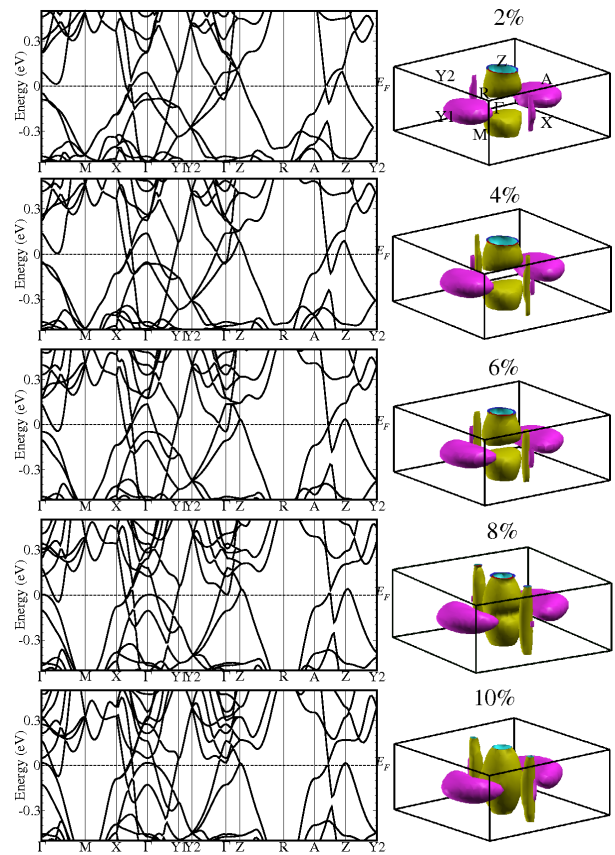}
\caption{\label{afmfs} (color online). Band structures (left column)
and Fermi surfaces (right column) of the striped antiferromagnetic
state with 2\%, 4\%, 6\%, 8\% and 10\% (from top to bottom)
compression of volume. The 4\%, 6\%, 8\% and 10\% corresponds to
pressures: 1.7 GPa, 3.1 GPa, 4.7 GPa, 6.4 GPa, and 8.3 GPa,
respectively. }
\end{figure}

In Fig. \ref{afmfs} we show the band structures and Fermi surfaces
with 2\%, 4\%, 6\%, 8\% and 10\% compression of volume. We can find
that the Fermi surfaces expand significantly due to the expansion of
band width by pressure. As volume compressed to 4\%, corresponding
to 3.1 GPa pressure where superconductivity has been observed, the
Fermi surfaces lie on antinodal directions of $\Gamma$ - Y1 line
expand and across $\Gamma$ - M line to invade neighbor quarters of
BZ. Thus, when pressure increases to 2.5 GPa where the
superconductivity takes place, the main change of Fermi surfaces of
the striped antiferromagnetic state is to expand in BZ. Furthermore,
as volume compressed over 8\%, the Fermi surfaces centered around Z
point across $\Gamma$ point to form a connected cylindrical tube.
The topological change of Fermi surfaces should respond to the slope
change of $c/a$ ratio about 8\% compression of volume (see Fig.
\ref{sum}). It is notable that the superconductivity disappears as
pressure is higher than 6 GPa, that is roughly same with the place
where the Fermi surfaces topological transition occurs.

%added by xie
However, it is necessary to point out that the shape of the Fermi
surfaces is dependent on the magnetic moment which is sensitive to
the exchange-correlation potential (LSDA or GGA) and the Fe-As bond
length, and thus different calculations give quite different Fermi
surfaces. Especially, the closed Fermi surfaces in the $\Gamma-Y$
line, which are mainly $d_{xy}$ character, seem not to be a common
feature of the iron-based superconducting family. The cylinder tubes
(see Fig. \ref{afmfs}), which are mainly $d_{xz/yz}$ character and
show two-dimensional behavior under pressure, are the common
features of the iron-based superconducting family. Although the
topological transition is observed under the pressure, such a
transition is sensitively dependent on the amplitude of magnetic
moment. As shown in Fig. \ref{sum}, the two-dimensional cylinder
tube in the $\Gamma-Z$ line appears when the moment is about 1.3
$\mu_{B}$ in the GGA calculation. Considering that the
experimentally observed magnetic moment is only about 0.8 $\mu_{B}$,
the cylinder-tube Fermi surfaces of the striped antiferromagnetic
state should appear at smaller pressure, because the current
electronic structure calculations always overestimate the magnetic
moment but describe the structural parameter very well. It should
give some clues  for the relationship between magnetism and
superconductivity, as discussed previously for LaFeAsO
system.\cite{lafeaso:pickett}

Finally, we investigate pressure effect on the band structures of
the nonmagnetic state. As discussed above, the Fe-As bond length is
always underestimated in the optimized structure of nonmagnetic
state under pressure, consequently $d_{xy}$ bands are below Fermi
level at $\Gamma$ point (See Fig. \ref{tetfs}), only shift downwards
a little with increasing pressure. While $d_{xz,yz}$ bands shift
downwards significantly under pressure, which across Fermi level as
compression of volume over 8\%, corresponding to a pressure value of
4.7 GPa in nonmagnetic calculations. Such a crossover will lead to a
Fermi surface topological transition of the hole-type cylinder-tube
closed at $\Gamma$ point. However, it should be mentioned that the
nonmagnetic results might not be true. Because ARPES experiments
always indicate the hole-type Fermi surfaces around $\Gamma$
connects with larger gap in the iron-pnictide materials family,
which should be important to
superconductivity\cite{arpes:ding,arpes:zhouxj}. Thus the
topological transition induced by pressure predicted by nonmagnetic
calculation would not coincide with the fact that $T_c$ reach
maximum nearby such a pressure.

It has been observed that the superconducting states coexist with
the striped magnetic state\cite{bafeas:staticmagsc}, but at present
we cannot say that the mechanism of the superconductivity is based
on the magnetism. It is just believed that the superconductivity in
the Fe based materials is unlikely mediated by phonons, but likely
originated from some magnetic or spin fluctuations. Our calculated
results indicate that the GGA calculations yield reasonable lattice
constants and $z_{\rm As}$ only when we consider the magnetism. This
implies that the magnetic effect is essential in these materials.
%added by xie 2009.01.20
Since the d$_{xz/yz}$ bands around Fermi level are determined by the
Fe-As antibonding, as the distance of Fe-As increases, the $k_z$
related hoppings are weakened and the bands become narrow,
the large experimental $z_{\rm As}$ parameter leads to typical
two-dimensional properties.
In current {\it ab initio} electron-phonon calculations, the
optimized structure of nonmagnetic state is used, which results in
the underestimation of the Fe-As bond length and more
three-dimensional character of Fermi surfaces, that would influence
properties of phonon and electron-phonon interaction. Such a
magnetism related effect might have substantial influence on the
electron-phonon coupling. A calculation including the magnetic
effect are highly expected to investigate how much the magnetic
background influences the electron-phonon interaction, to reveal the
possibility of that phonon might play some important role on the
superconductivity.
On the other hand, LDA calculations yield smaller magnetic moment
and underestimates the $z_{\rm As}$ parameter. The corresponding
Fermi surfaces is substantially different from that of ARPES
experiments\cite{arpes:feng}. If the superconductivity is not
directly dependent on the static moment, but determined by the
lattice parameters and spin fluctuations, the changing of the moment
value does not affect the superconductivity.

\section{compared with experimental results involved}

The main implication of pressure effect on the crystal structure of
BaFe$_2$As$_2$ is to shorten the Fe-Fe bond length and the Fe-As
bond length, but to maintain As-Fe-As bond angles simultaneously. We
could compare the pressure effect on crystal structure with the
doping effect in (Ba$_{1-x}$K$_x$)Fe$_2$As$_2$\cite{bakfeas:38K}.
With the doping of K into BaFe$_2$As$_2$, $T_c$ goes up and reaches
a maximum of 38 K at $x$=0.4. The impressive change is that As-Fe-As
angle $\varepsilon$ becomes the ideal tetrahedral angle of
109.5$^\circ$ as $x$=0.4. Under pressure, the FeAs$_4$ - tetrahedra
is hard to be changed and still far away from the ideal structure.
It seems to imply that superconductivity could not be improved
simply by hydrostatic pressure in BaFe$_2$As$_2$.

In the iron-pnictide superconducting family, the crystal structure,
magnetism and superconductivity play complex game. It has been
reported that a structural transition of $c$-axis collapse occurs in
CaFe$_2$As$_2$\cite{cafeas:instable,cafeas:yildirim} under pressure.
Such a structural phase transition happens nearby the upper boundary
of superconducting dome. In the striped antiferromagnetic
calculations of BaFe$_2$As$_2$, we find a collapse of $c$-axis as
the compression of volume above 8\%, which should be relevant to the
topological transition of Fermi surfaces. As the topological
transition of Fermi surfaces influence the magnetic moment of iron
slightly, the related $c/a$ collapse should not be relative to a
magnetic transition. In fact, we observe a magnetic collapse
accompanying by a sudden softness of $c/a$ ratio as lattice volume
is compressed over 16\%, corresponding to near 15 GPa pressure. It
is far away from the known superconducting dome (2-6 GPa) up to
date. We have performed LSDA calculations to examine the influence
of overestimated magnetic moment by GGA. We found that the LSDA
calculated magnetic moment is about 0.35 $\mu_B$ smaller than GGA,
and the magnetic collapse happens at about 14\% compression of
volume, corresponding to pressure about 13 GPa. These results are
consistent with recent experimental report that a suppression of
magnetic order by pressure is observed as pressure above about 13
GPa\cite{bafeas:13gpa}.

Now, we attempt to consider the implication of the striped
antiferromagnetic and nonmagnetic calculations for paramagnetic
state. It has been found experimentally that superconductivity
coexist with static magnetic order under pressure for BaFe$_2$As$_2$
\cite{bafeas:staticmagsc}, so that a strong magnetic background is
essential to superconductivity in these families. First of all, in
order to describe chemical bonding more accurately, the magnetic
effect should be considered. Since ferromagnetism is not stable
under a large range of pressure, only antiferromagnetic calculations
is possible. As the optimized internal parameter $z_{\rm As}$ of the
striped antiferromagnetic state is in good agreement with
experiments, it might be reasonable to use these parameters to
describe the crystal structures and electronic structures in
paramagnetic state than those nonmagnetic results. It coincides with
the fact that no significant structural transformation is observed
as system goes from antiferromagnetic state to superconducting state
or paramagnetic state, except the compression of $b/a$ orthorhombic
transition induced by the antiferromagnetic interaction. If the
optimized lattice parameters obtained from the striped
antiferromagnetic calculations are adopted, it turn out that
$d_{xz}$, $d_{yz}$ and $d_{xy}$ bands always cross Fermi level
except for shifting downwards a little under pressure. For such a
treatment of paramagnetic state, we found that the energy splitting
between $d_{xz,yz}$ and $d_{xy}$ bands decreases from 0.06 eV to
0.04 eV at $\Gamma$ point as volume is compressed 6\%, which
indicates that the $d_{xy}$ bands are determined by direct Fe-Fe
$d_{xy}$ interaction while the $d_{xz/yz}$ bands are mainly
determined by Fe $d_{xz,yz}$ - As $p_{x/y}$ interactions.
Nevertheless, the band shapes are maintained under pressure, and the
Fermi surfaces are quite similar to those calculations with
experimental lattice parameters (see Fermi surfaces of left column
in Fig. \ref{tetfs}), except the sheets around $\Gamma$ point expand
more. Thus the Fermi surfaces are robust against the pressure, no
topological transition is found as the lattice volume compressed up
to at least 10\%. The impressive features is very different from
those nonmagnetic calculations.

It should be reminded that the band structures near Fermi level, as
well as Fermi surfaces are sensitive to Fe-As bond length and
As-Fe-As angle, which highly depend on the details of the
spin-polarization. Our calculations indicate that the shape of bands
and Fermi surfaces are almost decided by the value of large range
magnetic moment. Therefore, if the pressure induces a high spin
polarized state transform to a low spin state (even nonmagnetic
state), not only $d_{xy}$ bands shift down to cross Fermi level, but
also $d_{xz,yz}$ bands cross Fermi level. This would certainly
influence the coupling between different Fermi surfaces, then
strongly influence the electronic properties. It would be very
interesting to examine the detail of magnetism under pressure by
experiment furthermore.

\section{conclusion }

In summary, we have employed an accurate all-electronic
full-potential linearized augmented plane wave method within the
density-functional theory to investigate the effect of pressure on
the crystal structure and electronic properties of BaFe$_2$As$_2$.
The calculated lattice parameters of the striped antiferromagnetic
state is in good agreement with experiment, while nonmagnetic state
calculation underestimates $c$-axis and Fe-As bond lengths and
yields a smaller $z_{\rm As}$. We find that the $c$-axis is
compressed more easily than $a$-axis in nonmagnetic calculation,
while the antiferromagnetic state calculation lead to a nearly
isotropic compressibility with pressure up to 6 GPa. With increasing
pressure, the FeAs$_4$ tetrahedra changes little. For the striped
antiferromagnetic state, we find that the Fermi surfaces get closer
and closer to the $\Gamma$ point in the $\pm z$ directions when
pressure increases from 2 GPa to 6 GPa, corresponding to the
pressure region where superconductivity occurs. We observe a Fermi
surface topological transition around $\Gamma$ point as the pressure
beyond 6 GPa. The transition should correspond to a small $c$-axis
collapse. A magnetic collapse inducing the softening of $c$-axis is
found when pressure reaches 13 GPa. These first-principles results
should be useful to understanding the antiferromagnetism and
electronic states in the FeAs-based materials, and may have some
useful implications to the superconductivity.

\acknowledgments

This work is supported  by Nature Science Foundation of China
(Grant Nos. 10704024, 10774180, and 10874232), by Shanghai
Rising-Star Program (Grant No. 08QA14026), and by the Chinese
Academy of Sciences (Grant No. KJCX2.YW.W09-5).

\end{document}